\documentclass[twocolumn,superscriptaddress,pre]{revtex4}
\usepackage{epsfig}
\usepackage{amssymb,amsfonts,amsmath}
\usepackage[pdftex]{pstricks}
\include{epsf}

\newcommand{\ben}{\begin{enumerate}}
\newcommand{\een}{\end{enumerate}}

\newcommand{\<}{\langle}
\renewcommand{\>}{\rangle}

\newcommand{\beq}{\begin{equation}}
\newcommand{\eeq}{\end{equation}}
\newcommand{\bea}{\begin{eqnarray}}
\newcommand{\eea}{\end{eqnarray}}

\usepackage{amsmath}
\usepackage{amssymb}
\usepackage{amsthm}
\usepackage{amsfonts}

\begin{document}
\title{Physical limit to concentration sensing amid spurious ligands}
\author{Thierry Mora}
\affiliation{Laboratoire de physique statistique, \'Ecole normale sup\'erieure, CNRS and UPMC, 24 rue Lhomond, 75005 Paris, France}

\date{\today}
\linespread{1}

\begin{abstract}
To adapt their
behaviour in changing environments, cells sense 
concentrations by binding external ligands to their receptors.
However, incorrect ligands may bind nonspecifically to receptors,
and when their
concentration is large, this binding activity may interfere
with the sensing of the ligand of interest. 
Here, I derive analytically the physical limit to the accuracy of
concentration sensing amid a large number of
interfering ligands. A scaling transition is found when the mean bound time of
correct ligands is twice that of incorrect ligands.
I discuss how the physical bound can be approached by a cascade of receptor
states generalizing kinetic
proof-reading schemes.
\end{abstract}

\maketitle

Because of their small sizes, biological systems typically operate
with only a few copies of the molecules they sense and communicate with.
In their pioneering work, Berg and Purcell derived the fundamental
bound that the noise arising
from these small numbers sets on the accuracy of concentration sensing \cite{Berg1977}. 
Experimental progress in the characterization of single-cell
variability \cite{Elowitz2002} and sensing precision \cite{Gregor2007}
has fueled a renewed interest in small-number noise and its implications for
information processing
\cite{Tkacik2011,Bowsher2014,Tkacik2014c}. General or refined bounds
on sensing accuracy
have been recently derived for single receptors
\cite{Bialek2005,Endres2009b,Kaizu2014}, and extended to spatial
\cite{Endres2008,Rappel2008,Rappel2008a,Endres2009a,Hu2010a} or temporal \cite{Mora2010a} gradient
sensing, while the metabolic cost and trade-offs of
sensing accuracy have been explored
\cite{Mehta2012,Lan2012,Becker2013,Lang2014,Govern2014,Govern2014a,Barato2014,Mancini2015,Barato2015,Hartich2015}.
Much of this past work has assumed perfect specificity
between the biological receptors and their cognate ligands. In
realistic biological contexts, large numbers of spurious ligands may
bind  receptors nonspecifically, interfering with the ligand of
interest \cite{Lalanne2015}. This is the case in the problem of
antigen recognition by T-cell receptors, where cells must react to a
small number of specific foreign peptides among a large number of
nonspecific self-peptides \cite{Feinerman2008}. Biochemical network
architectures based on kinetic proofreading
\cite{Hopfield1974,Ninio1975} have been shown to provide a solution to
the discrimination problem, and have been studied in depth theoretically
\cite{McKeithan1995,Francois2013,Lalanne2013,Lalanne2015}. 
However, no fundamental bound has been derived against which to compare
the performance of these solutions, save for
Ref.~\cite{Siggia2013} where concepts of statistical decision theory were
used to derive the minimal decision time to detect cognate
ligands. In this paper I derive the fundamental limit on concentration
sensing accuracy and ligand detection error in the presence of a large
number of spurious ligands. The maximum likelihood estimate achieving
the bound can be implemented biologically by simple networks based on push-pull reactions.

\begin{figure}
\begin{center}
\noindent\includegraphics[width=\linewidth]{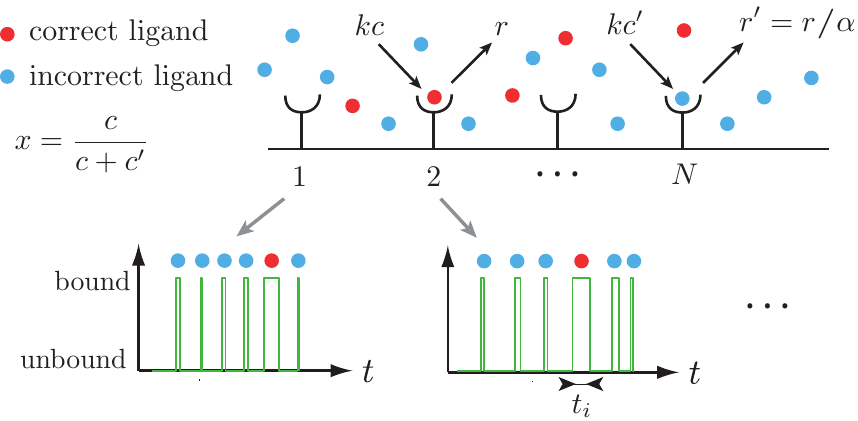}
\caption{{\bf Reading concentrations off trajectories of
    receptor occupancies.} Cognate (correct) and spurious (incorrect)
  ligands may bind $N$ receptors, typically presented on the cell
  surface, with rate $kc$ and $kc'$. The incorrect ligands are in
  excess, $c'>c$, but lead to shorter binding events, $r'>r$. The
  information the cell can theoretically use is contained in the time
  traces of occupancy of all receptors (green curves). The maximum
  likelihood estimate fully exploits these traces to optimally infer the input
  concentrations $c$ and $c'$.
\label{fig:measurer}
}
\end{center}
\end{figure}

Consider a mixture of two ligands, only one of which the biological system wishes
to sense. The ligand of interest (hereafter referred to as the correct ligand) is present in
concentration $c$, while the interfering or spurious ligand (called the incorrect ligand)
is present in concentration $c'$. The biological unit can sense 
ligands through $N$ identical receptors, which can be bound by either
ligand with a common rate $k=4Da$, where $D$ is the molecule
diffusivity and $a$ the effective receptor size.
Receptors can distinguish between the
two molecules thanks to their higher affinity to
the correct ligand. Physically, this means that the unbinding rate $r$ of the correct ligand
is smaller than that of the incorrect ligand $r'>r$.


The 
occupancy of each receptor,
\beq\label{p}
p=\frac{kcr^{-1}+kc' r'^{-1}}{1+kcr^{-1}+kc' r'^{-1}},
\eeq
depends on both concentrations, and cannot be used alone to determine
$c$. The interchangeability of the ratios $c/r$ and $c'/r'$ in this expression
emphasizes the ambiguity between many incorrect ligands and a few correct ones.
To discern these two effects, one must use the
full temporal record of occupancy of each receptor.
The probability distribution for the binding and unbinding events at
all receptors during a time interval $T$ reads:
\beq
P=e^{-kc_{\rm tot}T_{\rm u}}\prod_{i=1}^n \left( k c r
  e^{-r t_i} + k c' r' e^{-r' t_i}\right),
\eeq
with $c_{\rm tot}=c+c'$ is the total concentration of ligands, 
$T_u$ is the total unbound time accrued over all receptors, and $t_1,\ldots,t_n$ the durations of
the $n$ binding events occurring at all $N$ receptors during $T$.
The log-likelihod $\mathcal{L}=\ln P$ can be rewritten as a sum of three
independent contributions,
$\mathcal{L}=\mathcal{L}_0+\mathcal{L}_1+\mathcal{L}_2$, where
$\mathcal{L}_0$ depends on neither $c$ or $c'$, and where $\mathcal{L}_1$
and $\mathcal{L}_2$ pertain to the unbound and bound intervals respectively:
\bea
\mathcal{L}_1(c_{\rm tot})&=&n\ln c_{\rm tot} -kc_{\rm tot}T_{\rm u} \label{L1}\\
\mathcal{L}_2(x)&=&\sum_{i=1}^n \ln\left(1-x+x\alpha e^{(1-\alpha)r't_i}\right),
\eea
where $x=c/c_{\rm tot}$ is the fraction of correct ligands, and
$\alpha=r/r'<1$ is the binding constant ratio. As can be
seen in the respective
dependencies of $\mathcal{L}_1 $ and $\mathcal{L}_2$ upon $c_{\rm tot}$ and
$x$, unbound intervals are informative of
the total concentration, while the bound intervals are informative of the
fractions of ligands.

The maximum likelihood estimate for the total concentration is obtained by the condition $\partial
\mathcal{L}_1/\partial c_{\rm tot}=0$, which gives $c_{\rm
  tot}^*=kT_u/n$. The error made by this estimate is given
in the large time limit by the Cram\'er-Rao bound, which sets the best
possible performance of any estimator \cite{Cramer}:
\beq\label{CRctot}
\<\delta c_{\rm tot}^2\>\approx -{\left(\frac{\partial^2\mathcal{L}_1}{\partial c_{\rm tot}^2}\right)}^{-1}=\frac{c_{\rm tot}^2}{n}\approx\frac{c_{\rm tot}}{4Da (1-p)NT},
\eeq
with $\delta c_{\rm tot}=c_{\rm tot}^*-c_{\rm tot}$.
This result is that obtained in \cite{Endres2009b} for a single
ligand, where the maximum-likelihood error was shown
to be half as small as the classical Berg and Purcell bound
\cite{Berg1977} based on the average
receptor occupancy. The reason for this difference is that the maximum likelihood estimate is not
affected by the noise due to the stochastic nature of receptor unbinding, as
evident in Eq.~\eqref{L1}. In the case of a mixture, the receptor
occupancy Eq.~(\ref{p}) depends on $x$ as well as $c_{\rm tot}$, and does
not even suffice to determine the total concentration.

The fraction $x$ of correct ligands can be estimated by
maximum likelihood as well, by solving:
\beq\label{MLE}
\left.\frac{\partial
\mathcal{L}_2}{\partial x}\right|_{x^*}=
\sum_{i=1}^n\frac{\alpha e^{(1-\alpha)r't_i}-1}{1-x^*+x^*\alpha e^{(1-\alpha)r't_i}}=0.
\eeq
The error can be estimated from the Cram\'er-Rao bound (App.~A.1):
\beq\label{CR}
\begin{split}
\<\delta x^2\>&\approx -{\left(\frac{\partial^2\mathcal{L}_2}{\partial
      x^2}\right)}^{-1}\approx \frac{f(x,\alpha)}{n}\\
\textrm{with}\quad f(x,\alpha)^{-1}&=\int_0^{+\infty} du\, e^{-u} \frac{(\alpha
  e^{(1-\alpha)u}-1)^2}{1-x+x \alpha e^{(1-\alpha)u}}.
\end{split}
\eeq
The total error in the concentration of the correct
ligand $c=xc_{\rm tot}$ is then the sum of the (independent) errors in $c_{\rm tot}$ and $x$
from Eqs.~\eqref{CRctot} and \eqref{CR}: ${\<\delta c^2\>}\approx {c_{\rm tot}^2}{(x^2+f(x,\alpha))}/{n}$.

\begin{figure}
\begin{center}
\noindent\includegraphics[width=\linewidth]{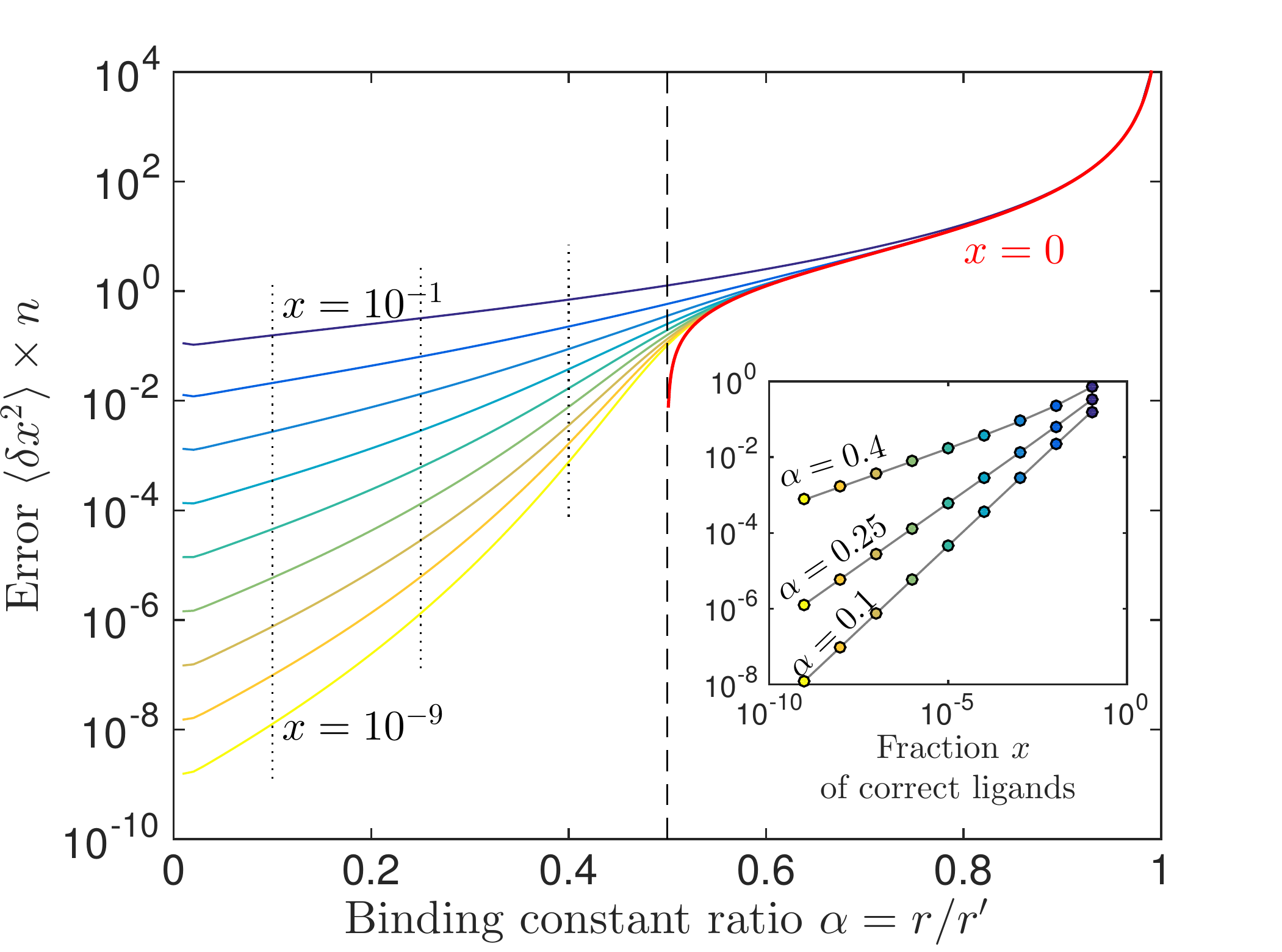}
\caption{{\bf Physical bound on concentration sensing error.} The
  fundamental bound on the relative error in the fraction of correct
  ligands $x=c/c_{\rm tot}$ scales with the inverse of the number of
  binding events $n$. Here the rescaled error $f(x,\alpha)=n\<\delta
  x^2\>$ [Eq.~\eqref{CR}], is represented as a function of the binding
  constant ratio $\alpha$ for various values of $x$. There are two distinct
  scaling regimes [Eq.~\eqref{smallx}]. For $\alpha>1/2$, the rescaled
  error depends weakly on $x$, while for $\alpha<1/2$ it scales as
  $x^\beta$, with $\beta=1-\alpha/(1-\alpha)$, as illustrated in the
  inset for three example values of $\alpha$.
\label{fig:error}
}
\end{center}
\end{figure}

It is interesting to consider the limit where the correct ligands are rare, $x\ll 1$, as in the case of immune recognition. Two scaling regimes, illustrated in Fig.~\ref{fig:error}, are found depending on the value of the ratio $\alpha$
between the two binding constants:
\beq\label{smallx}
f(x,\alpha)\approx\left\{\begin{array}{ll} g(\alpha)&
                                                                      \alpha>1/2,\\
h(\alpha) x^{\beta},\ \beta=1-\frac{\alpha}{1-\alpha}
                                                                    &\alpha<1/2,
                                                                      
\end{array}\right.
\eeq
with $g(\alpha)=(2\alpha -1)/(1-\alpha)^2$ and $h(\alpha)=(1-\alpha)\alpha^{-\frac{1}{1-\alpha}}
\sin(\pi\alpha/(1-\alpha))/{\pi}$, and $0<\beta<1$.
Since $x^2\ll f(x,\alpha)$, the error in $c$ reduces to:
\beq
\begin{split}
{\<\delta c^2\>}&\approx g(\alpha)\frac{c_{\rm tot}}{4Da(1-p)NT} \quad \alpha>1/2,\\
{\<\delta c^2\>}&\approx h(\alpha) \frac{c^{\beta}c_{\rm
  tot}^{1-\beta}}{4Da(1-p)NT}\quad\alpha<1/2.
\end{split}
\eeq
In the hard discrimination regime ($\alpha>1/2$), incorrect ligands
dominate the error, which is governed by $c_{\rm tot}$ as in Eq.~\eqref{CRctot}.
The prefactor  $g(\alpha)$ diverges at $\alpha= 1$, as expected when the two ligands have
the same binding constant and are thus indistinguishable. By contrast, in the easy discrimination regime ($\alpha<1/2$)
the error is governed by a weighted
geometric mean between $c$ and $c_{\rm tot}$.
In the limit of very small
$\alpha$, corresponding to clearly distinguishable ligands, the error is
$\<\delta c^2\>\approx c/[4Da(1-p)NT)]$---precisely the
error when no
interfering ligand is present \cite{Endres2009b}. For $x=0$, the
maximum-likelihood estimate may infer a small $x^*=\delta x>0$.
However, the second derivative of the likelihood diverges at $x=0$
for $\alpha<1/2$, indicating that
the Cram\'er-Rao bound \eqref{CR} fails to give a correct
estimate of this error, which instead scales anomalously with the
number of events: $\delta x\sim n^{\alpha -1}$, hence $\delta c\sim
c_{\rm tot}^{\alpha} [4Da(1-p)NT]^{\alpha-1}$ (App.~A.2).

\begin{figure}
\begin{center}
\noindent\includegraphics[width=.49\linewidth]{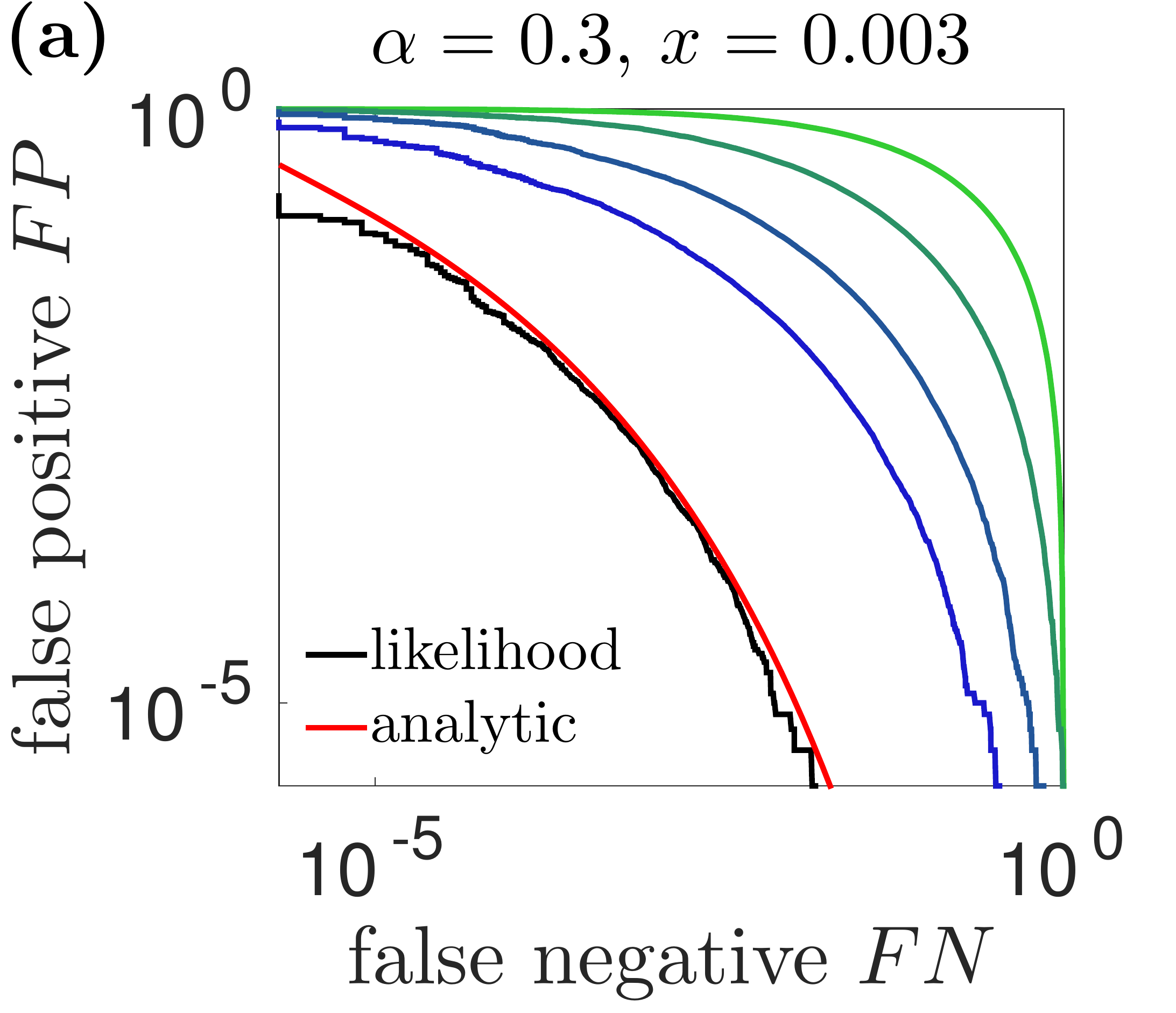}
\noindent\includegraphics[width=.49\linewidth]{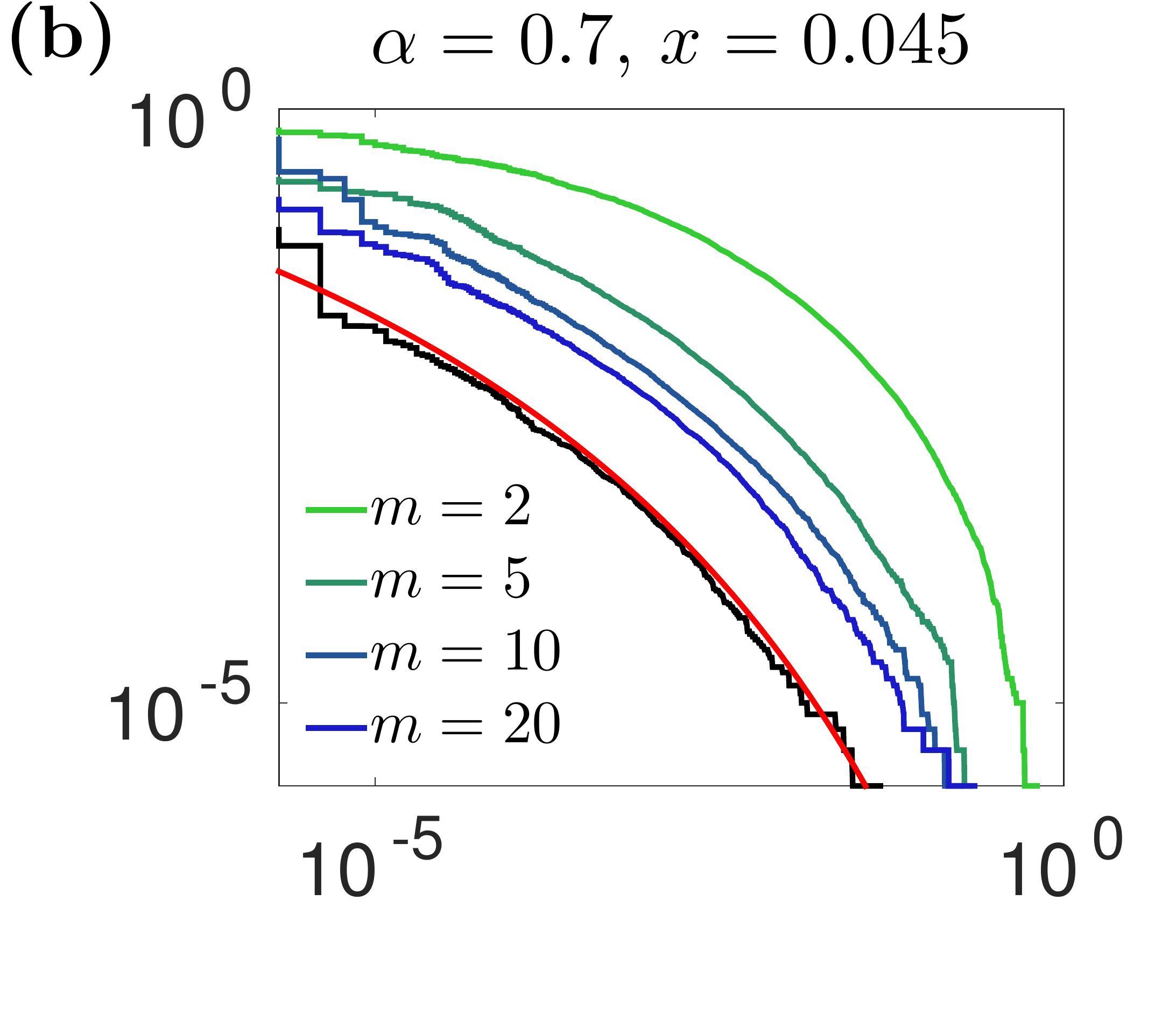}
\caption{{\bf Error in the detection of the correct ligand.} Numerical
  and analytical estimate of the rate of false positive (FP) versus
  false negative (FN) errors in the detection of a small fraction $x$
  of correct ligands, 
 for
  $n=10^5$ and (a) $\alpha=0.3$ and $x=0.003$ and (b) $\alpha=0.7$ and $x=0.045$. The black curve is the result of a numerical experiment, repeated $5\cdot 10^5$ times in presence of the correct ligand, and $5\cdot 10^5$ without, and where a likelihood ratio test $\mathcal{L}_2(x)>\theta$ was used with a varying threshold $\theta$. The red curve is the analytical prediction from Eq.~\eqref{roc}. The green to blue curves show the performance of optimized networks schematized in Fig.~\ref{fig:network}, for various numbers of receptor states $m$.
\label{fig:roc}
}
\end{center}
\end{figure}

In many situations, it is more useful for the system to determine the
presence of the correct ligand rather than its precise
concentration, as in the recognition of
foreign pathogens by immune receptors. This decision can be made
optimally (in the Bayesian sense) by comparing
the likelihoods of the two competing hypotheses: presence {\em versus}
absence of the correct ligand at fraction $x$. The
presence of the correct ligand is detected when
$\ln[P(\{t_i\}|x)/P(\{t_i\}|0)]=\mathcal{L}_2(x)>\theta$, where $\theta$ is an
adjustable parameter controlling the balance between the
  false-positive and false-negative error rates $FP$ and $FN$. These errors decay
  exponentially fast with large numbers $n$ of binding events, and can be
  estimated in that limit using a saddle-point approximation (App.~B1):
\beq
\label{roc}
FP\approx \frac{\exp\left[n\phi(\lambda)-\lambda\theta\right]}{\lambda\sqrt{2\pi
    n|\phi''(\lambda)|}},\quad 
FN\approx \frac{\lambda}{1-\lambda}e^\theta FP,
\eeq
where $\phi(\lambda)=\ln\left[\int_0^{+\infty}du\, e^{-u}[1-x+\alpha x
e^{(1-\alpha)u}]^\lambda\right]$, and where $\lambda$ satisfies the
saddle-point condition $\theta=n\phi'(\lambda)$. The receiver operating
characteristics (ROC) giving the dependency between $FP$ and $FN$ can thus be
estimated parametrically by varying $\lambda$. This saddle-point approximation is well verified by
numerical simulations (Fig.~\ref{fig:roc}).

As in the case of concentration sensing error, a scaling transition is found in the limit of scarce correct ligands, $x\ll 1$.
When $\alpha>1/2$, one obtains
\beq
\label{roc2}
FP \approx \frac{e^{-\frac{1}{2} \lambda^2n
    x^2/g(\alpha)}}{\sqrt{2\pi\lambda^2n x^2/g(\alpha)}},\ 
FN \approx \frac{e^{-\frac{1}{2} (1-\lambda)^2n x^2/g(\alpha)}}{\sqrt{2\pi
    (1-\lambda)^2n x^2/g(\alpha)}},
\eeq
while when $\alpha<1/2$ both error rates decay as $\sim (n
x^\gamma)^{-1/2}\exp[-C n x^\gamma]$, with
$\gamma=(1-\alpha)^{-1}$, $1<\gamma<2$ and $C$ a function of $\alpha$
and $\lambda$ (App.~B.2).
The
time $T$ necessary to make a reliable decision scales as
$[4Da(1-p)N]^{-1}c_{\rm tot}  c^{-2}$ for $\alpha>1/2$, and as
$[4Da(1-p)N]^{-1}c_{\rm tot}^{\gamma-1}c^{-\gamma}$ for $\alpha <1/2$.
Equivalent scaling laws were obtained in \cite{Siggia2013}
for minimal on-the-fly detection times.

Can biological systems approach the physical bound on concentration
sensing given by Eq.~\eqref{CR}? To gain insight into this question,
one can expand Eq.~\eqref{MLE} at first order in $x$ to get an
approximation to the maximum likelihood estimate when $\alpha>1/2$
(for $\alpha<1/2$ this expansion gives quantities with
 diverging means and cannot be used):
\beq\label{MLEapprox}
x^*\approx  \frac{2\alpha-1}{(1-\alpha)^2}\frac{1}{n}\sum_{i=1}^n \left(\alpha e^{(1-\alpha)r't_i}-1\right).
\eeq
This estimator, which is subject to the same asymptotic error as in
Eq.~\eqref{smallx}, suggests a simple strategy, where each receptor
signals ``positively'' with a rate that depends on how long it has been
bound, $\alpha (1-\alpha) r' e^{(1-\alpha)r't}$, and ``negatively'' ({\em
  i.e.} with an opposite effect on the readout, see below) through
a fixed burst $(\alpha-1)\delta(t)$ upon binding, so that the net
effect of each binding event $i$ on the readout molecule concentration is
\beq\label{scheme}
\int_{0}^{t_i}dt \left[\alpha (1-\alpha) r' e^{(1-\alpha)r't}\right] +\alpha-1=\alpha e^{(1-\alpha)r't_i}-1,
\eeq
{\em i.e.} exactly the argument of the sum in Eq.~\eqref{MLEapprox}.

\begin{figure}
\begin{center}
\noindent\includegraphics[width=\linewidth]{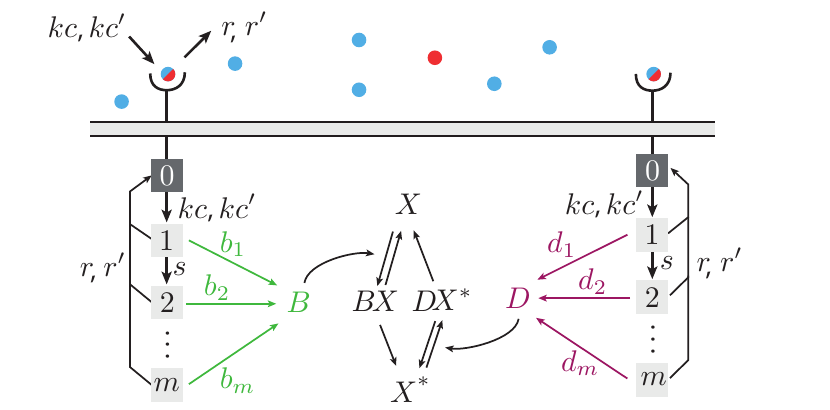}
\caption{{\bf Network for sensing the concentration of correct
    ligands.} Upon binding, each receptor enters a cascade of $m$
  states along which it proceeds with rate $s$. The ligand can detach
  at any moment with rate $r$ or $r'$ depending on its identity
  (correct or incorrect), bringing the receptor back to the unbound
  state 0. While in conformational state $\mu=1,\ldots,m$, the
  receptor activates two enzymes $B$ and $D$ with rates $b_\mu$ and
  $d_\mu$, each catalyzing two oppposite Michaelis-Mentens reactions
  of a pull-push network. $B$ and $D$ are assumed to equilibrate fast
  and to be always limiting in the reactions they catalyze, so that
  $dX^{*}/dt \propto b_{\mu}-d_{\mu}$.
\label{fig:network}
}
\end{center}
\end{figure}

This idea can be
implemented biologically by a cascade of receptor conformational
states triggered by binding, and proceeding irreversibly from states
$1$ to $m$, each transition to the next state occurring with rate
$s$ (Fig.~\ref{fig:network}). The ligand is free to detach from the
receptor at any time, bringing the receptor back to the unbound state
$0$. The receptors signal through the production or activation of two
molecules $B$ and $D$ with opposite effects on a push-pull network
governing the state of a molecule $X$, which provides the final
readout for $x$ through its modified state $X^*$. If one 
requires that the equilibration of $B$ and $D$ are fast, and that $X$ and $X^*$ are
always in excess in the Michaelis-Mentens reactions, then
\beq
\frac{dX^*}{dt}=X_0\sum_{j=1}^N(b_{\mu(j)}-d_{\mu(j)}), \quad X_0=\textrm{const,}
\eeq
where $\mu(j)$ is the state of the $j^{\rm th}$ receptor, and $b_0=d_0=0$.
For the purpose of this discussion, the internal molecules $B$, $D$
and $X$ are assumed to be unaffected by biochemical noise, restricting
the source of noise to the input alone. In this design $X^*$ increases indefinitely to
  mimick the sum in Eq.~\eqref{MLEapprox} over all events at all
  receptors. A more a realistic but equivalent scheme would involve a
  running sum over an effective time $T$, obtained by
  relaxing $X^*$ to $X$ with rate $\sim T^{-1}$ \cite{Mora2010a}.

When the number of states $m$ is large and the transitions between
them are rapid,  $X^*$ can
  track Eq.~\eqref{MLEapprox} with arbitrary precision when $\alpha>1/2$.
In that case, the receptor state $\mu$ provides an approximation to
the time since binding, $\mu\approx st$. Then, for
  example, receptors signaling
positively with rate $b_\mu \propto \alpha (1-\alpha) r'
e^{(1-\alpha)r'\mu/s} \approx\alpha (1-\alpha) r'
e^{(1-\alpha)r't}$, and negatively with rate $d_\mu \propto (1-\alpha)
(s/\mu_0)e^{-\mu/\mu_0}\approx (1-\alpha)\delta(t)$ (with $\mu_0$ an adjustable
  parameter) would exactly realize
Eq.~\eqref{scheme} and thus the estimator of
Eq.~\eqref{MLEapprox} in the limit $m\gg s/r\gg \mu_0\gg 1$.


\begin{figure}
\begin{center}
\noindent\includegraphics[width=.49\linewidth]{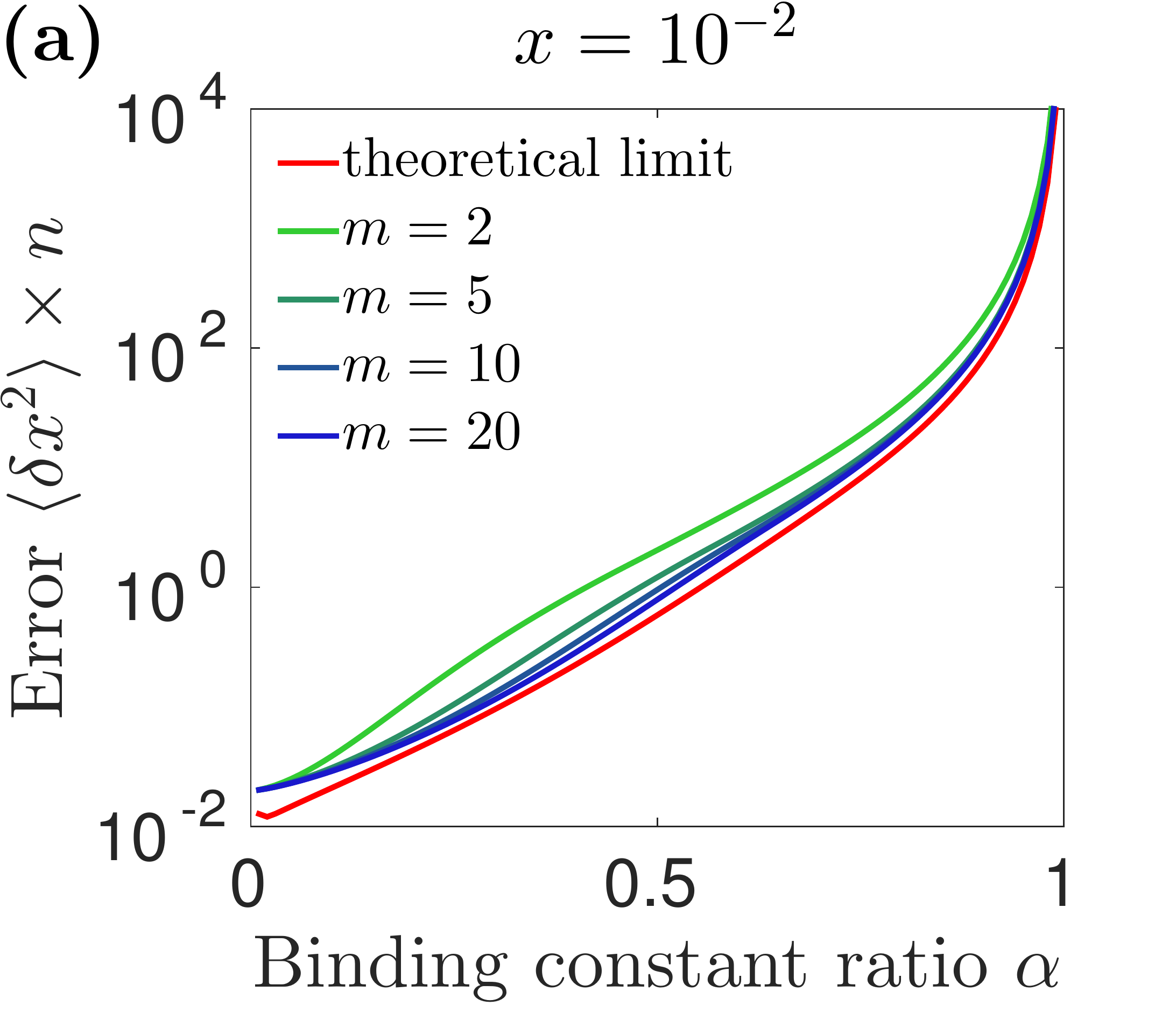}
\noindent\includegraphics[width=.49\linewidth]{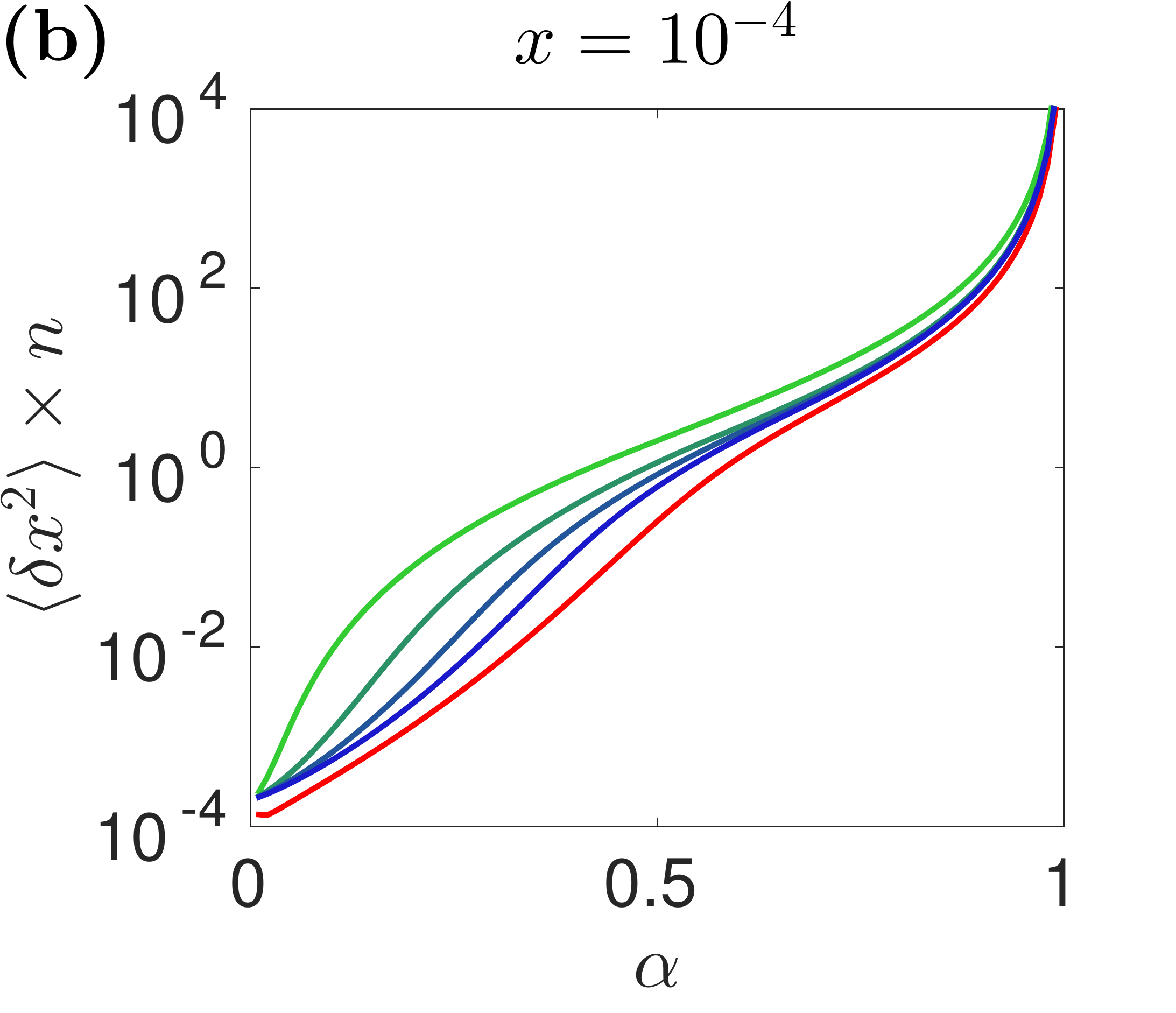}
\caption{{\bf Network performance.} Error made by optimized networks with a finite number of receptor states $m$ (green to blue curves), compared to the theoretical bound (red curve), for (a) $x=10^{-2}$ and (b) $x=10^{-4}$.
\label{fig:netperf}
}
\end{center}
\end{figure}

Although such optimal performance is only reached for large
$m$ and $\alpha>1/2$, this network design may still perform
well in more general situations. 
One can
optimize the expected error produced by this network over the net
signaling rates $(b_\mu-d_\mu)$, with the 
constraint that the mean effect of binding incorrect ligands on
$X^*$ be zero, so that $\Delta X^* \propto c$ on average (App.~C).
Fig.~\eqref{fig:netperf} shows how the performance of such
optimized networks approaches the theoretical bound as the number of
states $m$ increases. The convergence is significantly worse for 
$\alpha<1/2$ at small $x$. In that regime, the estimator of \eqref{MLEapprox}
is not valid, suggesting that this network design may not achieve the
optimal bound even with an infinite number of states. The output of
these networks can also be used to detect ligands. Their performance
in doing so is compared to the optimal discrimination errors of
Eq.~\eqref{roc} in Fig.~\eqref{fig:roc}.

The principle of maximum likelihood not only yields the fundamental
bound on the accuracy of discerning cognate ligands from spurious ones, but also
suggests biochemical solutions to approach this optimal bound. Such
maximum-likelihood inspired designs have been previously
proposed in the case of a single ligand
\cite{Mora2010a,Lang2014}.
The network structure proposed in this study (Fig.~\ref{fig:network}) is reminiscent of kinetic proofreading schemes
and their generalizations, which provide a well-known solution to the
ligand discrimination problem
\cite{Hopfield1974,Ninio1975,Murugan2012a,Lalanne2013,Lalanne2015}. An
important difference is that here signaling occurs during all steps,
albeit at various, fine-tuned rates, and with potentially negative contributions, the
role of which is to buffer the effect of wrong ligands. Consistent
with this prediction, it was shown
that a negative interaction through a diffusible molecule between
kinetic-proofreading receptors could
mitigate the effects of large numbers of incorrect ligands
in a discrimination task \cite{Lalanne2015}.

The present results are relevant beyond the particular case of sensing by
receptors, and apply to any kind of biochemical signaling in
presence of
competing ligands or ``cross-talk.'' This is the case for example in
the context of gene
regulation, where competing transcription factors may bind regulatory
sites unspecifically, a problem particularly acute in metazoans \cite{Cepeda2015}.

The scaling transition occurring at the binding constant ratio $\alpha=1/2$
suggests that different strategies should be employed depending on how
hard the discrimination task is. In particular, the approximate but
biologically implementable
estimator of Eq.~\eqref{MLEapprox} curiously breaks down in the easy discrimation
regime, $\alpha<1/2$.
In that regime, the optimal bound is harder to achieve because it is
dominated by rare, long binding events that are hard to encode by
biochemical solutions.
The example of
immune recognition falls precisely into that regime, with a
binding constant ratio $\alpha$ between agonist and nonagonist ligands
ranging from one fifth to one third
\cite{Feinerman2008}. More elaborate network designs, probably with
feedback, may be needed to achieve the theoretical bound Eq.~\eqref{CR} in that
case. Finally, this study has assumed throughout that the
unbinding rates $r$ and $r'$ are priorly known to the system. Complex
mixtures of ligands with unknown 
binding constants would make for interesting generalizations.

I thank A. Walczak for her helpful comments on the manuscript.
While this article was under review, a paper treating a
  similar topic was submitted to the arXiv \cite{Singh2015}.

\appendix



\section{Cram\'er-Rao bound}

\subsection{The Cram\'er-Rao bound is tight: a physicist's proof}
In general the Cram\'er-Rao bound is a lower bound on the error made by any unbiased estimator, but it is not always certain whether this bound can be achieved.
Here the maximum likelihood estimate is shown to approach the
Cram\'er-Rao bound in the limit of large samples.

Assume that the likelihood of the data factorizes over independent
datapoints,
\beq
\mathcal{L}=\ln P=\sum_{i=1}^n \ell(x,t_i),
\eeq
where $x$ is the model parameter to be estimated, and
$(t_1,\ldots,t_n)$ the series of datapoints. In the specific case of receptors
binding to two types of ligands, $x$ is the fraction of correct
ligands, $t_i$ the duration of binding event $i$, and
\beq\label{siLL}
\ell(x,t_i)=\ln r' - r't_i+\ln\left(1-x+x\alpha
e^{(1-\alpha)r't_i}\right).
\eeq
The derivative of $\ell$ with respect to $x$ is denoted by
$\ell'(x,t_i)=\partial \ell(x,t_i)/\partial x$.
The maximum likelihood estimate $x^*$ satisfies:
\beq
\sum_{i=1}^n \ell'(x^*,t_i)=0.
\eeq
This estimator is unbiased: if $\tilde x$ denotes the true parameter
with which the data was generated, then $x^*$ should give back $\tilde x$ on average. Equivalently, 
\beq
\begin{split}
\left\<\frac{\partial\mathcal{L}(\tilde x)}{\partial x}\right\>_{\tilde x}=&\sum_{i=1}^n \<\ell'(\tilde x,t_i)\>_{\tilde x}= n \int_0^{+\infty} dt\, e^{\ell(\tilde x,t)} \ell'(\tilde x,t_i) \\
=& n\left.\frac{\partial}{\partial x}\int_0^{+\infty} dt\, e^{\ell(x,t)}\right|_{\tilde x}=0,
\end{split}
\eeq
(the last integral is just 1 because of normalization), where
$\<\cdot\>_{\tilde x}$ denote averages over data generated with the true parameter
$\tilde x$. In other words, the maximum of $\mathcal{L}$ is reached at $\tilde x$ on average. The probability that this maximum $x^*$ be larger than a certain value $x>\tilde x$ is:
\beq
\mathbb{P}(x^*>x)=\mathbb{P}\left(\frac{\partial\mathcal{L}(x)}{\partial x}>0\right)=\left\<\Theta\left(\frac{\partial\mathcal{L}}{\partial x}\right)\right\>_{\tilde x}.
\eeq
The Heaviside function $\Theta$ can be replaced by its Fourier representation:
\beq\label{siTheta}
\Theta(x) =\int_{-\infty}^{+\infty} \frac{d\omega}{2\pi \omega}e^{i\omega x}=\int_{-i\infty}^{+i\infty} \frac{d\lambda}{2\pi i \lambda}
e^{\lambda x},
\eeq
allowing for factorization over datapoints:
\beq
\begin{split}\label{siPcum}
\mathbb{P}(x^*>x)&=\int_{-i\infty}^{+i\infty} \frac{d\lambda}{2\pi i
  \lambda} \prod_{i=1}^n \int_0^{+\infty} dt_i e^{\ell(\tilde
  x,t_i)+\lambda \ell'(x,t_i)}\\
&=\int_{-i\infty}^{+i\infty} \frac{d\lambda}{2\pi i
  \lambda} \exp\left[n\ln \int_0^{+\infty} dt e^{\ell(\tilde
  x,t)+\lambda \ell'(x,t)}\right].
\end{split}
\eeq
and 
\beq
\begin{split}
&P(x)=-\frac{d \mathbb{P}(x^*>x)}{d x}=\\
&=-\int_{-i\infty}^{+i\infty} \frac{d\lambda}{2\pi i}
  \left[\int_0^{+\infty} dt \ell''(x,t) e^{\ell(\tilde
  x,t)+\lambda \ell'(x,t)}\right]\\
&\quad\times\exp\left[(n-1)\ln \int_0^{+\infty} dt e^{\ell(\tilde
  x,t)+\lambda \ell'(x,t)}\right].
\end{split}
\eeq

This integral can be evaluated by a saddle-point approximation in the
large $n$ limit:
\beq
\int d\lambda G(\lambda)e^{nF(\lambda)} \approx
G(\lambda^*)\sqrt{\frac{2\pi}{n|F''(\lambda^*)|}} e^{nF(\lambda^*)},
\eeq
with
\begin{align}
G(\lambda)&=\frac{\int_0^{+\infty} dt \ell''(x,t) e^{\ell(\tilde
  x,t)+\lambda \ell'(x,t)}}{\int_0^{+\infty} dt e^{\ell(\tilde
  x,t)+\lambda \ell'(x,t)}},\\
F(\lambda)&=\ln \int_0^{+\infty} dt e^{\ell(\tilde
  x,t)+\lambda \ell'(x,t)}.
\end{align}

The
saddle $\lambda^*$ is given by the condition that the derivative of the argument
of the exponential with respect to $\lambda$ be zero:
\beq\label{sisaddle}
\int_0^{+\infty} dt \ell'(x,t) e^{\ell(\tilde
  x,t)+\lambda \ell'(x,t)}=0.
\eeq
At $x=\tilde x$, this condition is satisfied for $\lambda=0$.
In the limit of large samples, $x^*-\tilde x$
is small and so should the corresponding $\lambda$.
One can expand at small $x-\tilde x>0$
and $\lambda$: 
\beq\label{siexp}
\ell'(x,t)\approx \ell'(\tilde x,t)+(x-\tilde x)\ell''(\tilde
x,t),
\eeq
and
\beq
\int_0^{+\infty} dt \left[(x-\tilde x) \ell''(\tilde x,t)+\lambda
  \ell'(\tilde x,t)^2 \right] e^{\ell(\tilde
  x,t)}=0,
\eeq
yielding $\lambda=x-\tilde x$ and:
\beq
P(x)\approx \frac{\sqrt{nH}}{\sqrt{2\pi}}
\exp\left[-\frac{n}{2} (x-\tilde x)^2 H\right],
\eeq
with
\beq
H=-\int_0^{+\infty}dt \ell''(\tilde x,t)
  e^{\ell(\tilde x,t)}=\int_0^{+\infty}dt \ell'(\tilde x,t)^2
  e^{\ell(\tilde x,t)}.
\eeq
A symmetric argument gives the same result for $x<\tilde
x$. The resulting distribution of $x^*$ is Gaussian, with mean
$\tilde x$ and variance
\beq
\<\delta x^2\>=\frac{1}{nH}=-{\left\<\frac{\partial^2
      \mathcal{L}(\tilde x)}{\partial x^2}\right\>_{\tilde x}}^{-1}={\left\<\left(\frac{\partial
      \mathcal{L}(\tilde x)}{\partial x}\right)^2\right\>_{\tilde x}}^{-1}.
\eeq

In the specific case of Eq.~\eqref{siLL},
\beq
H=\int_0^{+\infty} dt r'e^{-r't}\frac{(\alpha e^{(1-\alpha)r't}-1)^2}{1+x(\alpha e^{(1-\alpha)r't}-1)}.
\eeq
Performing the change of variable $u=r't$ yields the result of the
main text:
\beq\label{siCR}
\begin{split}
\<\delta x^2\>&\approx \frac{f(x,\alpha)}{n}\\
\textrm{with}\quad f(x,\alpha)^{-1}&=\int_0^{+\infty} du\, e^{-u} \frac{(\alpha
  e^{(1-\alpha)u}-1)^2}{1-x+x \alpha e^{(1-\alpha)u}}.
\end{split}
\eeq

\subsection{Small $x$ limit}
For $\alpha>1/2$ and small $x$, the integral in Eq.~\eqref{siCR} can
be approximated by:
\beq\label{sihard}
f(x,\alpha)^{-1}\approx \int_0^{+\infty} du\, e^{-u} (\alpha
  e^{(1-\alpha)u}-1)^2=\frac{(1-\alpha)^2}{2\alpha -1}.
\eeq

For $\alpha <1/2$, the function $e^{-u} (\alpha
  e^{(1-\alpha)u}-1)^2$ is not integrable, and the denominator of
  Eq.~\eqref{siCR} is necessary to ensure integrability at large $t$, however small
  $x$ is. Thus, the values of $u$ governing the behavior of the integral
 satisfy $x\alpha e^{(1-\alpha)u}=O(1)$. This observation
  suggests the change of variable $y=x\alpha e^{(1-\alpha)u}$:
\beq\label{sif}
f(x,\alpha)^{-1}=x^{-\beta} \alpha^{\frac{1}{1-\alpha}}\frac{1}{1-\alpha} \int_{\alpha x}^{+\infty} dy\,
  y^{-\frac{2-\alpha}{1-\alpha}} \frac{(y-x)^2}{1+y-x},
\eeq
with $\beta=1-\alpha/(1-\alpha)$. Expanding $(y-x)^2$ gives three terms
scaling as $x^{-\beta}y^{-\alpha/(1-\alpha)}$, $x^{1-\beta}y^{-1/(1-\alpha)}$ and
$x^{2-\beta}y^{-\frac{2-\alpha}{1-\alpha}}$ at small $y$, respectively. The last
two give diverging integrals as $x\to 0$ for all $\alpha$, 
yielding terms of order 1. Only when $\alpha>1/2$ does the first term give a diverging
integral, and thus a term of order 1 in $x$; in that case, the sum of all three terms gives back
the result of Eq.~\eqref{sihard}. If $\alpha<1/2$ however, the first term
is integrable and thus dominates the expression for $x\ll 1$, yielding:
\beq
f(x,\alpha)^{-1}=x^{-\beta} \alpha^{\frac{1}{1-\alpha}}\frac{1}{1-\alpha} \int_{0}^{+\infty} dy\,
  \frac{y^{-\frac{\alpha}{1-\alpha}}}{1+y}+O(1),
\eeq
where $O(1)$ denotes a term of order 1 at small $x$.
The integral can be calculated:
\beq
\int_{0}^{+\infty} dy\,
  \frac{y^{-\frac{\alpha}{1-\alpha}}}{1+y}=\frac{\pi}{\sin\left(\pi\frac{\alpha}{1-\alpha}\right)},
\eeq
to finally obtain:
\beq
f(x,\alpha)^{-1}\approx x^{-\beta} \alpha^{\frac{1}{1-\alpha}}\frac{1}{1-\alpha} \frac{\pi}{\sin\left(\pi\frac{\alpha}{1-\alpha}\right)}.
\eeq

In the intermediate case $\alpha=1/2$, the three terms in the integral
of Eq.~\eqref{sif} are of order $y^{-1}$, $xy^{-2}$  and
$x^2y^{-3}$. Again the last two terms diverge in the integral and give
contributions of order 1. The first term also diverges, but its
contribution reads:
\beq
\alpha^{\frac{1}{1-\alpha}}\frac{1}{1-\alpha} (-\ln(\alpha x)) =
\frac{1}{2}|\ln(x/2)|,
\eeq
so that:
\beq
f(x,\alpha)^{-1}=\frac{1}{2}|\ln(x/2)| + O(1).
\eeq

When $x=0$ and $\alpha\leq 1/2$, $f(x,\alpha)=\infty$, as the large
deviation function of $x^*$ becomes nonanalytic. The expansion
of $\ell'$ in Eq.~\eqref{siexp}  is no longer integrable when done around $\tilde
x=0$, and needs revisiting. The integral in
Eq.~\eqref{siPcum} reads:
\begin{widetext}
\beq
\int_0^{+\infty} du e^{-u}\exp\left[\frac{\lambda(\alpha
  e^{(1-\alpha)u}-1)}{1+x(\alpha e^{(1-\alpha)u}-1)}\right]\\
=1+\int_0^{+\infty} du e^{-u}\left\{\exp\left[\frac{\lambda(\alpha
  e^{(1-\alpha)u}-1)}{1+x(\alpha e^{(1-\alpha)u}-1)}\right]-1-\lambda(\alpha
  e^{(1-\alpha)u}-1)\right\}
\eeq
where the same change of variable $u=r't$ has been done. Doing a
further change of variable to $y= \alpha x e^{(1-\alpha)u}$ yields:
\beq\label{siint}
1+\frac{(\alpha x)^{\frac{1}{1-\alpha}}}{1-\alpha} \int_{\alpha
  x}^{+\infty} y^{-\frac{2-\alpha}{1-\alpha}} \left\{\exp\left[\frac{(\lambda/x) (y -x)}{1+y-x}\right]-1-(\lambda/x) (y -x)\right\}
\eeq
\end{widetext}
The term is the brackets is of order $(y-x)^2$, as was the case in
Eq.~\eqref{siCR}. Hence, terms in $y^2$ are integrable and dominate
the expression, which becomes at leading order in $x,
\lambda$:
\beq
1+\frac{(\alpha x)^{\frac{1}{1-\alpha}}}{1-\alpha} \int_{0}^{+\infty} y^{-\frac{2-\alpha}{1-\alpha}} \left\{\exp\left[\frac{(\lambda/x) y}{1+y}\right]-1-(\lambda/x) y\right\}
\eeq
With $\tilde\lambda=\lambda x$, the saddle-point condition becomes:
\beq
\psi'(\tilde \lambda,\alpha)=0,
\eeq
with
\beq
\psi(\tilde \lambda, \alpha)=-\frac{\alpha ^{\frac{1}{1-\alpha}}}{1-\alpha}\int_{0}^{+\infty} y^{-\frac{2-\alpha}{1-\alpha}}
\left(e^{\frac{\tilde\lambda y}{1+y}}-1-\tilde \lambda  y\right)
\eeq
and the cumulative probability distribution is:
\beq\label{sisaddle2}
\mathbb{P}(x^*>x)\approx \frac{1}{\sqrt{2\pi nx^{\frac{1}{1-\alpha}}\psi''(\tilde \lambda,\alpha)} \tilde \lambda}
  \exp\left[-nx^{\frac{1}{1-\alpha}}\psi(\tilde \lambda, \alpha)\right],
\eeq
where $\psi'=\partial\psi/\partial \tilde \lambda$ and $\psi''=\partial^2 \psi/\partial \tilde \lambda^2$. Fluctuation of $x^*$ are thus of order $n^{\alpha-1}\ll 1/\sqrt{n}$.

When $\alpha=1/2$, the term of order $(y-x)^2$ in the brackets of Eq.~\eqref{siint} dominates and diverges, so that this expression reduces at leading order to:
\beq
1+\frac{x^2}{2}|\ln(x/2)| \left(\frac{\tilde \lambda^2}{2}-\tilde \lambda\right).
\eeq
The saddle point condition gives $\tilde \lambda =1$ and one obtains:
\beq
\mathbb{P}(x^*>x)\approx \frac{1}{\sqrt{\pi nx^{2}|\ln(x/2)|}}
  \exp\left[-\frac{nx^{2}|\ln(x/2)|}{4}\right],
\eeq
which implies fluctuations of order $\delta x\sim (n\ln n)^{-1/2}$.

\section{Probability of discrimination error}

\subsection{General case}

The discrimination between two competing hypotheses---presence  {\em versus} absence of the
correct ligand in fraction $x$---can be performed by a
likelihood ratio test:
\beq
\ln\frac{P(t_1,\ldots,t_n|x)}{P(t_1,\ldots,t_n|x=0)} =\sum_{i=1}^n \left[\ell(x,t_i)-\ell(0,t)\right]>\theta,
\eeq
where $\theta$ is an adjustable parameter.
The false-positive and false-negative error rates are defined
as the probabilities of detecting the presence of a ligand that is in
fact absent,
and of missing it where it is there:
\beq
\begin{split}
FP&=\int \prod_{i=1}^n [dt_i e^{\ell(0,t_i)}] \Theta\left(\sum_{i=1}^n
  \left[\ell(x,t_i)-\ell(0,t)\right]-\theta\right),\\
FN&=\int \prod_{i=1}^n [dt_i e^{\ell(x,t_i)}] \Theta\left(\sum_{i=1}^n
  \left[\ell(0,t_i)-\ell(x,t)\right]+\theta\right).
\end{split}
\eeq
The integral representation of the Heaviside function, Eq.~\eqref{siTheta} can
be used again to obtain:
\beq
\begin{split}
FP&=\int_{-i\infty}^{+i\infty} \frac{d\lambda}{2\pi i \lambda} e^{-\lambda
  \theta} {\left[\int dt e^{(1-\lambda)\ell(0,t)+\lambda
    \ell(x,t)}\right]}^n,\\
FN&=\int_{-i\infty}^{+i\infty} \frac{d\lambda}{2\pi i \lambda} e^{\lambda
  \theta} {\left[\int dt e^{(1-\lambda)\ell(x,t)+\lambda
    \ell(0,t)}\right]}^n.
\end{split}
\eeq
Substituting $\lambda\to (1-\lambda)$ in the second equation gives an
expression for $FN$ that looks very similar to $FP$:
\beq
\int \frac{d\lambda}{2\pi i (1-\lambda)} e^{(1-\lambda)
  \theta} {\left[\int dt e^{(1-\lambda)\ell(0,t)+\lambda.
    \ell(x,t)}\right]}^n.
\eeq
In summary:
\beq
\begin{split}
FP&=\int_{-i\infty}^{+i\infty} \frac{d\lambda}{2\pi i \lambda} e^{n\phi(\lambda) -\lambda
  \theta}\\
FN&=\int_{-i\infty}^{+i\infty} \frac{d\lambda}{2\pi i (1-\lambda)} e^{n\phi(\lambda) +(1-\lambda)
  \theta},
\end{split}
\eeq
with:
\beq
\begin{split}
\phi(\lambda)&=\ln \int dt e^{(1-\lambda)\ell(0,t)+\lambda \ell(x,t)}\\
& = \ln \int du\, e^{-u} \left[1+x(\alpha e^{(1-\alpha)u}-1)\right]^{\lambda}.
\end{split}
\eeq
These two expressions can be evaluated in the large $n$
limit using a saddle-point approximation, with the same saddle-point
condition $\theta=n\phi'(\lambda)$ for both, yielding:
\beq
\begin{split}\label{roc3}
FP&\approx \frac{1}{\lambda\sqrt{2\pi
    n|\phi''(\lambda)|}}\exp\left[n\phi(\lambda)-\lambda\theta\right],\\
FN&\approx \frac{1}{(1-\lambda)\sqrt{2\pi
    n|\phi''(\lambda)|}}\exp\left[n\phi(\lambda)+(1-\lambda)\theta\right].
\end{split}
\eeq

\subsection{Small $x$ limit}
Again two regimes emerge in the $x\ll 1$ limit, depending on whether $\alpha$ is smaller or greater than $1/2$.
When $\alpha>1/2$, $\phi(\lambda)$ can be expand at small $x$:
\beq
\phi(\lambda)\approx -\frac{1}{2}\frac{(1-\alpha)^2}{2\alpha-1} \lambda(1-\lambda) x^2.
\eeq
This implies:
\beq
\label{siroc2}
\begin{split}
FP &\approx \frac{e^{-\frac{1}{2} \lambda^2n
    x^2/g(\alpha)}}{\sqrt{2\pi\lambda^2n x^2/g(\alpha)}},\\
FN &\approx \frac{e^{-\frac{1}{2} (1-\lambda)^2n x^2/g(\alpha)}}{\sqrt{2\pi
    (1-\lambda)^2n x^2/g(\alpha)}},
\end{split}
\eeq
with
\beq
g(\alpha)=\frac{2\alpha-1}{(1-\alpha)^2}.
\eeq

When $\alpha<1/2$, one can do the same change of variable as before,
$y=x\alpha e^{(1-\alpha)u}$, to obtain at leading order:
\beq
\phi(\lambda)=\frac{(\alpha
x)^{\frac{1}{1-\alpha}}}{1-\alpha}\int_0^{+\infty}dy\,
y^{-\frac{2-\alpha}{1-\alpha}}\left[(1+y)^\lambda -1-\lambda y \right].
\eeq
The integrand is of order $y^{-\alpha/(1-\alpha)}$ at small $y$, and
therefore is integrable. The error rates are then given by:
\beq\label{siroc3}
\begin{split}
FP&\approx \frac{\exp\left[nx^{\frac{1}{1-\alpha}}\left(\chi(\lambda)-\lambda\chi'(\lambda)\right)\right]}{\lambda\sqrt{2\pi
    nx^{\frac{1}{1-\alpha}}|\chi''(\lambda)|}}\\
FN&\approx \frac{\exp\left[nx^{\frac{1}{1-\alpha}}\left(\chi(\lambda)+(1-\lambda)\chi'(\lambda)\right)\right]}{(1-\lambda)\sqrt{2\pi
    nx^{\frac{1}{1-\alpha}}|\chi''(\lambda)|}}.
\end{split}
\eeq
where
\beq
\chi(\lambda)=\frac{\alpha^{\frac{1}{1-\alpha}}}{1-\alpha}\int_0^{+\infty}dy\,
y^{-\frac{2-\alpha}{1-\alpha}}\left[(1+y)^\lambda-1-\lambda y\right].
\eeq

The intermediate case $\alpha=1/2$ is treated similarly as before, by
noting that the integral defining $\phi(\lambda)$ is dominated by the
(diverging) term of order $y^{-1}$. This gives:
\beq
\phi(\lambda)\approx -\frac{x^2}{4} |\ln(x/2)| \lambda(1-\lambda).
\eeq
and therefore:
\beq
\label{siroc4}
\begin{split}
FP &\approx \frac{e^{-\frac{1}{4} \lambda^2 n
    x^2|\ln(x/2)|}}{\sqrt{\pi\lambda^2n x^2|\ln(x/2)|}},\\
FN &\approx \frac{e^{-\frac{1}{4} (1-\lambda)^2n x^2|\ln(x/2)|}}{\sqrt{\pi
    (1-\lambda)^2n x^2|\ln(x/2)|}}.
\end{split}
\eeq

As a result, the number of binding events $n$ necessary to a make reliable decision scales as $x^{-2}$ for
$\alpha>1/2$, $x^{-2}|\ln(x)|^{-1}$ for $\alpha=1/2$ and $x^{-\gamma}$
for $\alpha<1/2$, with $\gamma=(1-\alpha)^{-1}$. Replacing $n\approx
4Da c_{\rm tot} (1-p)NT$ gives the scaling for the minimal detection time:
\beq
T\sim\frac{1}{4Da(1-p)N}\times\left\{\begin{array}{ll}
c_{\rm tot}  c^{-2}&\alpha>1/2,\\
c_{\rm tot} c^{-2} |\ln(c/c_{\rm tot})|^{-1} & \alpha=1/2,\\
c_{\rm tot}^{\gamma-1}c^{-\gamma} & \alpha <1/2.
\end{array}\right.
\eeq

\section{Optimization of the signaling rates in the receptor cascade}
Each receptor goes through a cascade of states $\mu=1,\ldots,m$ upon
binding. At any moment, the receptor can become unbound with rate
$r$ or $r'$. In the following some expressions will be given in terms of the
unbinding rate of the correct ligand $r$, but the same expressions hold
for the incorrect ligand after substitution by $r'$.

The probability of reaching state $\mu$ is
$[s/(s+r)]^{\mu-1}$. Assuming it has reached state $\mu$, the time
$t_{\mu}$ spent in that state is distributed according to
$(s+r)e^{-(s+r)t_{\mu}}$. In summary $t_{\mu}$ is distributed as
follows:
\beq
P_r(t_{\mu})=\frac{s^{\mu-1}}{(s+r)^{\mu-2}} e^{-(s+r)t_{\mu}}
+ \left[1-{\left(\frac{s}{s+r}\right)}^{\mu-1}\right] \delta(t_{\mu}),
\eeq
where $\delta(x)$ is Dirac's delta function.
Its first and second moments are:
\begin{align}
\<t_\mu\>_r&=\frac{s^{\mu-1}}{(s+r)^{\mu}},\\
\<\delta t_\mu^2\>_r&=\frac{s^{\mu-1}}{(s+r)^{\mu+1}} \left[2-\frac{s^{\mu-1}}{(s+r)^{\mu-1}}\right].
\end{align}
The output of the network is given by:
\beq
\frac{dX^*}{dt}=X_0\sum_{j=1}^N (b_{\mu(j)}-d_{\mu(j)}),
\eeq
where $\mu(j)$ is the state of the $j^{\rm th}$ receptor and $b_0-d_0=0$,
so that the net effect of one binding event is
\beq
\Delta X^*=  X_0\sum_{\mu=1}^m (b_\mu-d_\mu) t_\mu.
\eeq
On average, binding a wrong ligand will cause a change
\beq
\<\Delta X^*\>_{r'}= X_0\sum_{\mu=1}^m (b_\mu-d_\mu) \frac{s^{\mu-1}}{(s+r')^\mu}.
\eeq
When optimizing over the net rates $(b_\mu-d_\mu)$, this quantity is
set to zero, to ensure that only the correct ligand changes
$X^*$ on average. This way, $X^*$ is proportional to $c$ in the limit of
long times:
\beq
\<X^*(T)\> \approx 4Da(1-p) NT c \<\Delta X^*\>_r,
\eeq
Although the mean of $X^*(T)$ is not
affected by incorrect binding events, its variance is, and reads:
\beq
\begin{split}
&\<X^*(T)^2\>-\<X^*(T)\>^2 \approx 4Da(1-p) NT c_{\rm tot}  \\
&\ \times \left[x\<(\Delta X^*)^2\>_r
+ (1-x)\<(\Delta X^*)^2\>_{r'} - x^2 \<\Delta X^*\>_r^2\right],
\end{split}
\eeq
where
\beq
\<(\Delta X^*)^2\>_{r}=\<\Delta X^*\>_{r}^2 +  X_0^2\sum_{\mu=1}^m
(b_\mu-d_\mu)^2 \<\delta t_\mu^2\>_r.
\eeq
and the same for $r'$.

For a given $m$, the signal-to-noise ratio
\beq
SNR=\frac{\<X^*(T)\>^2}{\<X^*(T)^2\>-\<X^*(T)\>^2}
\eeq
is maximized over the rates $b_\mu-d_\mu$. The procedure gives the
optimized networks discussed in the main text.

\bibliographystyle{pnas}
\bibliography{rightfromwrong,other}

\end{document}